\documentclass[conference]{IEEEtran}

\usepackage{times}
\usepackage{latexsym}
\usepackage{graphicx}
\usepackage{adjustbox}
\usepackage{float}
\usepackage{epstopdf}
\usepackage{multirow}

\usepackage{hyperref}

\usepackage{soul,color}
\usepackage[justification=centering]{caption}

\usepackage{multicol}

\begin{document}

\title{Enhancing Efficiency and Privacy in Memory-Based Malware Classification through Feature Selection}


\author{\IEEEauthorblockN{Salim Sazzed}
\IEEEauthorblockA{\textit{Department of Computer Science}}
\IEEEauthorblockA{Old Dominion University, Norfolk, VA, USA \\ University of Memphis, Memphis, TN, USA} \\
\and
\IEEEauthorblockN{Sharif Ullah}
\IEEEauthorblockA{\textit{Department of Computer Science and Engineering}} 
\IEEEauthorblockA{University of Central Arkansas, Conway, AR, USA}
}




\maketitle

\begin{abstract}
Malware poses a significant security risk to individuals, organizations, and critical infrastructure by compromising systems and data. Leveraging memory dumps that offer snapshots of computer memory can aid the analysis and detection of malicious content, including malware. To improve the efficacy and address privacy concerns in malware classification systems, feature selection can play a critical role as it is capable of identifying the most relevant features, thus, minimizing the amount of data fed to classifiers. In this study, we employ three feature selection approaches to identify significant features from memory content and use them with a diverse set of classifiers to enhance the performance and privacy of the classification task. Comprehensive experiments are conducted across three levels of malware classification tasks: i) binary-level benign or malware classification, ii) malware type classification (including Trojan horse, ransomware, and spyware), and iii) malware family classification within each family (with varying numbers of classes). Results demonstrate that the feature selection strategy, incorporating mutual information and other methods, enhances classifier performance for all tasks. Notably, selecting only 25\% and 50\% of input features using Mutual Information and then employing the Random Forest classifier yields the best results. Our findings reinforce the importance of feature selection for malware classification and provide valuable insights for identifying appropriate approaches. By advancing the effectiveness and privacy of malware classification systems, this research contributes to safeguarding against security threats posed by malicious software.
\end{abstract}

\section{Introduction}
Malware, an abbreviation for malicious software, encompasses software or code specifically designed to inflict harm or compromise computer systems, networks, or users. It encompasses various forms, including viruses, worms, Trojans, ransomware, spyware, and adware \cite{carrier2022detecting}. The proliferation of physical communication systems such as smartphones, tablets, Internet of Things (IoT) devices, and cloud computing has led to a surge in the development and deployment of malware \cite{demme2013feasibility}. For instance, global malware attacks reached 5.5 billion in 2022, indicating a two percent increase compared to the previous year \footnote{Source: \url{https://www.statista.com/statistics/873097/malware-attacks-per-year-worldwide/}}. It should be noted that malware attacks are not only a threat to the privacy and security of individuals' or organizations' data, but they may also extend to critical infrastructures and become fatal to human lives. For example, the Triton malware attack (2017), a state-sponsored attack, targeted a petrochemical plant to take over the safety instrument systems of the plant. The aim was to kill humans by triggering an explosion or releasing toxic gas \cite{Malware:Critical}. Besides, the healthcare and public health sectors are also increasingly being affected by malware attacks; in 2022 alone, a total of 210 ransomware incidents were reported in these sectors \cite{Malware:Healthcare}.


As mentioned before, malware can have different forms and exploit particular sectors of targets. To determine the level of risk and severity associated with different malware, the types or families of malware need to be identified and prioritized for proactive and reactive cyber defense. Analyzing the traits of malware plays a crucial role in understanding, identifying, and effectively countering the threats they present. Specifically, categorizing different types of malware is of utmost importance as it helps in crafting appropriate responses, attributing attacks to their sources, and developing proactive security measures to safeguard systems, networks, and users from potentially harmful consequences.

Malware analysis can be categorized into three types: static, dynamic, and memory-based \cite{sihwail2019malware}. In static analysis, malicious files are studied without being executed, and the required features are extracted accordingly. As there is no need for execution, static methods require much fewer computational resources along with providing a fast recognition scheme. However, recent malware files use obfuscation techniques, such as the insertion of dead code, register reassignment, the substitution of instruction, and code manipulation to avoid static analysis
detection \cite{ye2017survey}. In contrast, behavior analysis executes and monitors malicious files in a controlled environment. Unlike static analysis, behavior analysis is not vulnerable to obfuscation techniques, but it consumes excessive time and resources (i.e. memory and CPU usage) \cite{cheng2017shellcode}.

Memory analysis has been proven to be a powerful analysis technique that can effectively study malware behaviors \cite{rathnayaka2017efficient}. It uses memory images to analyze information about running programs, operating systems, and the general state of the computer. Examining memory can detect process/DLL hooking techniques used by malware to appear as a legitimate process. The analysis provides accurate information about malware behaviors by extracting memory-based features that can express malware activities and characteristics. Memory-based features can also overcome some of the behavior analysis limitations, such as the single view of execution and malware’s disguised behaviors during the analysis \cite{dai2018malware}. Dissecting memory dumps is an effective approach for detecting and classifying malware \cite{lashkari2021volmemlyzer}, in addition to other approaches such as static and dynamic analysis \cite{sihwail2019malware}. Given the sophisticated evasion techniques employed by malware to circumvent traditional security measures, examining the memory of infected systems becomes essential for gaining crucial insights. Memory analysis unveils hidden processes, injected code, or rootkit-like behaviors that may escape detection by antivirus or intrusion detection systems. Exploring the memory space occupied by running processes makes it possible to determine their objectives, functionalities, and potentially malicious actions. This valuable information contributes to enhancing cybersecurity defenses by strengthening existing security measures.

Malware variants are continuing to evolve by using advanced obfuscation and packing techniques. These concealing techniques make malware detection and classification significantly challenging. Manually scrutinizing memory dumps to comprehend compromised processes, network connections, and code artifacts associated with malware is time-consuming and requires significant effort \cite{abbasi2022behavior}. Apart from accuracy, timely malware detection is crucial to minimize the potential damage and impact caused by the malware. Swift identification of malware allows for prompt actions like isolating infected systems, removing the malware, and restoring compromised data or configurations. Due to this urgency, researchers have conducted numerous studies in recent years focusing on automatic malware identification, including their existence and more detailed analysis, such as type classification. These studies make use of various machine learning techniques to achieve their objectives. By leveraging machine learning, researchers aim to expedite the detection process and enhance the efficiency of malware analysis, thus bolstering overall cybersecurity efforts.

Feature selection can play an important role in classification and prediction \cite{tsafrir2023efficient}. However, they remain mostly unexplored in malware detection \cite{abbasi2022behavior}. A carefully curated feature set can improve the performance of classifiers, as shown in earlier studies from diverse domains \cite{lehavi2022feature}. Although, in most areas, the primary purpose of feature selection is to reduce computational complexity and enhance accuracy, in cyber-security domains,  another crucial aspect is privacy, which can be benefited from feature selection. As data privacy is a critical component in the cyber-security domain, limiting the amount of data utilized for training machine learning models can greatly reduce the risk of security breaches.

Therefore, here, we focus on identifying influential features from memory dumps to limit the data used in the classification task. Since improper feature selection ( e.g., erroneous exclusion of essential features, introduction of bias and errors) can lead to performance degradation, we explore multiple feature selection approaches. As in the cyber-security domain, computational time is an important element, we focus solely on filter-based feature selection methods, which are computationally efficient and do not require exhaustive search. We consider three filter-based feature selection methods- i) Chi-square, ii) Analysis of variance (ANOVA), and iii) Mutual information (MI) for limiting the number of features utilized for classification. To validate the advantage of the feature selection, we conduct comprehensive experiments across three levels of malware classification tasks: i) binary-level benign or malware classification, ii) malware type classification (Trojan horse, ransomware, spyware), and iii) malware family classification for each malware family (5 classes based on malware family) using selected features with multiple classifiers. Our results demonstrate that feature selection strategies enhance the performance of classifiers across all tasks. Notably, by selecting only 25\% and 50\% of input features, when using random forest (RF) classifiers, we achieve similar or better results than incorporating all the features (100\%). MI feature selection approach with Random Forest (RF) classifier shows the most consistent performance.

\subsection{Contributions}
The primary contributions of this paper can be summarized as follows-

\begin{itemize}
 \item We investigate the effectiveness of diverse feature selection approaches for malware identification and classification task.

\item We demonstrate that the MI-based feature selection approach can effectively identify significant memory features (e.g., 25\%, 50\%) and can obtain a similar or better performance than utilizing all the input features.  
\end{itemize}

\section{Related Work}
In recent years, there has been a significant increase in malware attacks, leading to a growing focus on malware analysis as a critical research area in the cybersecurity domain. Malware leaves distinct traces in computer memories, making memory analysis a valuable tool to gain insights into malware behavior and patterns. As our study revolves around malware detection through memory content, this section primarily delves into works related to memory analysis.

Dener et al. \cite{dener2022malware} applied a number of machine learning algorithms for malware and benign type classification (2-class) from memory data (the same dataset considered in this study). The authors obtained accuracy close to 100\% for this malware identification task. The authors of the MalMemAnalysis-2022 dataset \cite{carrier2022detecting} utilized an ensemble approach for distinguishing benign and malware classes from memory data and obtained an  F1 score of around 0.99.

A number of studies incorporated image processing algorithms for malware identification and classification tasks. For example, Dai et al. \cite{dai2018malware} presented a method that involves extracting a memory dump file and converting it into a grayscale image. The author(s) resized the image to a fixed size,  extracted features using the histogram of gradients technique, and then employed a classification algorithm to categorize the malware. Li et al. (2019) proposed a deep learning-based approach for malware analysis. This method involves taking a memory snapshot of the system and converting it into a grayscale image. They employed a convolutional neural network (CNN) to model the system and train the deep learning model to distinguish between malicious and benign memory snapshots. The authors claimed that this approach significantly reduces analysis runtime without compromising accuracy. Bozkir et al. \cite{bozkir2021catch} proposed a new memory dumping and computer vision based method to detect malware in memory, even they do not exist on the hard drive. The proposed approach captures the memory dump of suspicious processes which are converted to RGB images. The authors then generate feature vectors utilizing GIST and HOG (Histogram of Gradients) descriptors as well as their combination, which are then classified by machine learning classifiers.

Mosli et al. \cite{mosli2016automated} conducted a study to detect malware by extracting Registry, DLLs, and APIs from memory images which compared malware detection performances using machine learning algorithms. Later the authors performed a behavior-based automated malware detection using forensic memory analysis and machine learning techniques \cite{mosli2017behavior}. Petrik et al. \cite{petrik2018towards} performed malware detection more specifically with binary raw data from memory dumps of devices. It had the characteristics of being independent from the operating system and architectural structure. Demme et al. \cite{demme2013feasibility} demonstrated the effectiveness of leveraging various performance counters, such as instructions per cycle (IPC), cache behavior, and memory behavior, to classify malware. Building upon this research, Tang et al. \cite{tang2014unsupervised} utilized hardware performance counters (HPC) in conjunction with unsupervised methods to detect malware. Sharafaldin et al. \cite{sharafaldin2017botviz} proposed BotViz, a hybrid method that incorporates hooks to enhance bot detection. Martin-Perez et al. \cite{martin2021pre} introduced two strategies (Guided De-Relocation and linear Sweep De-Relocation) for pre-processing memory dumps, aiming to expedite and simplify the analysis process by relocating file objects. 

Only a limited number of works are available which incorporate feature selection for malware classification. Abbasi et al. \cite{abbasi2022behavior} proposed a particle swarm optimization (PSO) based meta-heuristic approach for feature selection. However, their work focused on only ransomware detection. Besides, they employed the costly wrapper-based feature selection approach, which is computationally expensive. Moreover, they focused on malware behavior-based analysis, which is fundamentally different from our memory-based analysis. Tsafrir et al. \cite{tsafrir2023efficient} introduced three feature extraction methodologies for MP4 malware detection and incorporated them with machine learning (ML) algorithms. These methodologies included two file structure-based approaches and one knowledge-based approach. To assess their effectiveness, the researchers conducted a series of experiments using six ML algorithms on multiple datasets.

Some other works focused on creating tools for memory analysis and forensics. For example, Okolica and Peterson \cite{okolica2010compiled} introduced CMAT, a self-contained tool that can extract forensic information from the memory dump. The authors emphasized the significance of a highly flexible memory analysis process that can be employed across different platforms and systems in their study. Such flexibility can significantly reduce the time required to match the system with the corresponding profile. Block and Dewald \cite{block2017linux} introduced a memory analysis plug-in, which was designed to simplify the analysis process. The plug-in provides detailed information about heap objects in memory and can aid memory analysis professionals in understanding operations occurring within the system memory. Lashkari et al. \cite{lashkari2021volmemlyzer} developed VolMemLyzer, a Python-based script designed to facilitate feature extraction from memory dumps generated by another tool called Volatility. VolMemLyzer extracts thirty-six features from memory dumps, encompassing various categories, such as processes, dynamic link libraries, sockets, handles, callbacks, loaded modules, code injections, and connections. 

\begin{figure*}[t]
\includegraphics[width=0.6 \linewidth]{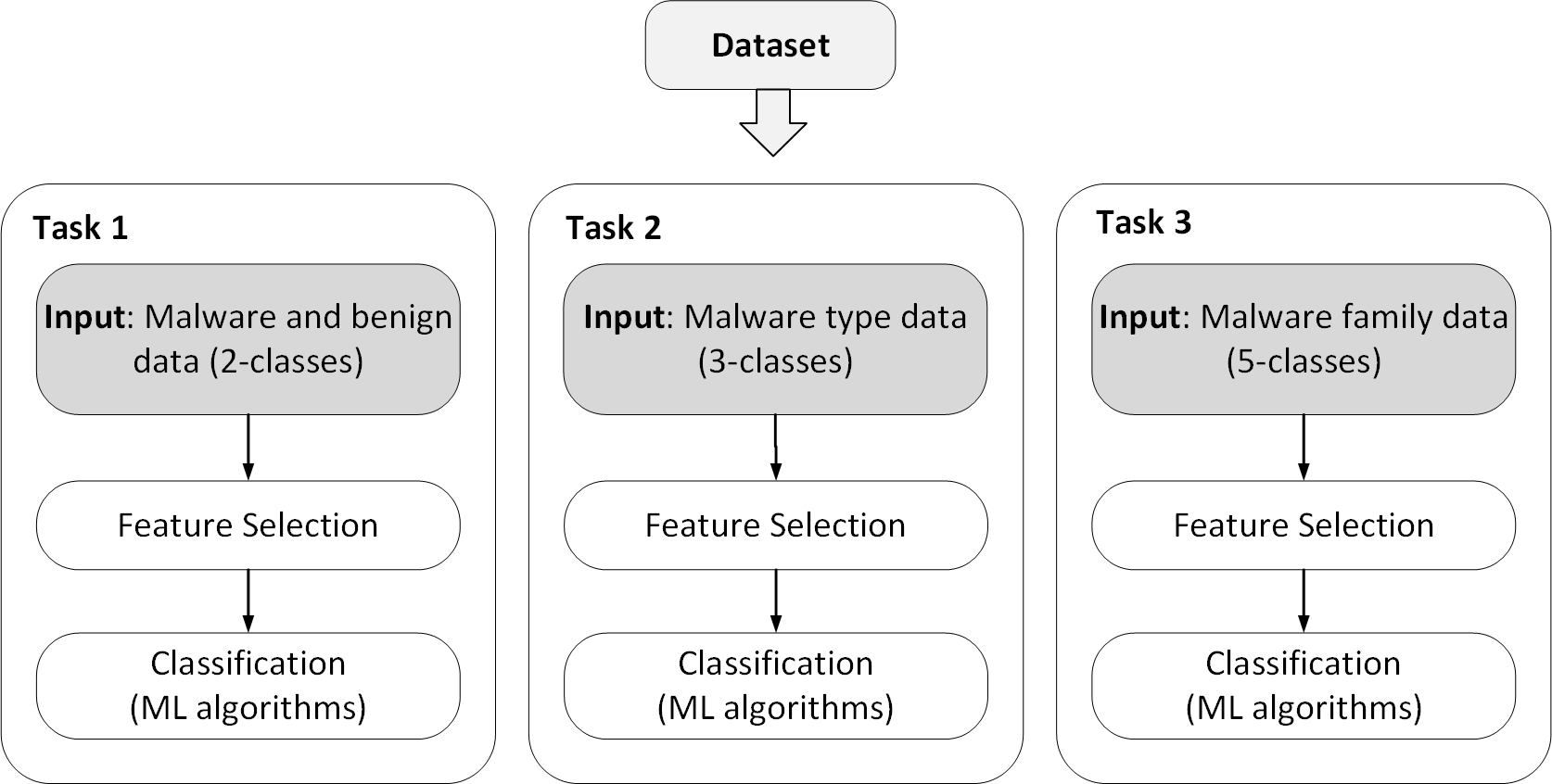}
\vspace{3mm}
\centering \caption{Various levels of classifcation tasks considered in this study}
\label{classification_task}
\end{figure*}

\section{Dataset}

The MalMemAnalysis-2022 dataset considered in this study was introduced by Carrier et al. \cite{carrier2022detecting}, which includes hidden malware families obtained by memory analysis. The dataset consists of 58,596 memory records, equally divided into benign and malicious (i.e., malware) classes, each having 29,298 instances. There are 56 features in the MalMemAnalysis-2022  dataset, with each feature representing specific types of memory information (more details regarding dataset creation and features are available on \cite{carrier2022detecting}).

The 29298 malware samples comprise the following three types of malware-

\begin{enumerate}
    \item Trojan horse: A Trojan horse is a type of malicious software (malware) that disguises itself as a legitimate program or file to deceive users into installing or executing it on their computers. Once inside, it can perform harmful actions without the user's knowledge, such as stealing sensitive data, providing unauthorized access to cybercriminals, recording keystrokes, participating in botnets, and delivering ransomware. Trojans are often distributed through deceptive means like email attachments or infected downloads, posing significant risks to computer security. 

    \item Spyware: Spyware is a type of malware designed to covertly gather information from a user's computer or device without their knowledge or consent. Once installed, spyware operates in the background, tracking browsing habits, recording keystrokes, capturing login credentials, and monitoring other sensitive data. The collected information is then sent back to the attacker, who can use it for various malicious purposes, such as identity theft, financial fraud, or targeted advertising. Spyware's focus is on information gathering and stealthy monitoring of the user's activities. It typically does not have destructive capabilities like a Trojan.

    \item Ransomware: Ransomware is a type of malicious software (malware) designed to encrypt the files and data on a victim's computer or network, rendering them inaccessible. Once the files are encrypted, the ransomware displays a ransom note, demanding payment, usually in cryptocurrency, in exchange for providing the decryption key needed to unlock the files. 
    It is typically distributed through malicious email attachments, infected software downloads, or by exploiting vulnerabilities in systems. 
\end{enumerate}

\begin{table}[!ht]
\caption{Malware Statistics in  MalMemAnalysis-2022 dataset}
\centering
\begin{tabular}{c|c|c}
\textbf{Malware Type (\#Total)} & \textbf{Malware Family} & \textbf{(\#)Instances}  
\\
& Zeu & 1950 \\
& Emotet & 1967 \\
Trojan Horse (9487)  &  Refroso & 2000 \\
& scar & 2000 \\
& Reconyc & 1570 \\
\hline

& 180Solutions  &2000 \\
& Coolwebsearch & 2000 \\
Spyware (10020) & Gator & 2200\\
& Transponder & 2410 \\
& TIBS & 1410\\
\hline
& Conti & 1988 \\
& MAZE & 1958\\
Ransomware (9791) & Pysa & 1717 \\
& ako & 2000  \\
& Shade & 2128 \\
\end{tabular}
\label{dataset_statitics}
\end{table}

Each type of malware further can be categorized into multiple malware families, as shown in detail in Table \ref{dataset_statitics}.

Table \ref{dataset_statitics} shows the distributions of malware types and families in the dataset. As mentioned earlier, three types of malware are present in the dataset, where each type further contains malware from five different families. Considering the 3-class malware types (i.e., Trojan horse, Spyware, and Ransomware ),  the dataset is almost class-balanced. The dataset representing each type of malware, based on the distributions of corresponding malware families,  is not entirely class-balanced,  as among the five malware families, the percentage of a malware family can range from 15\% to 25\%.

\section{Malware Categorization}

Here, we utilize memory dumps for malware detection/classification at three distinct levels (Fig \ref{classification_task}), where each level is associated with a particular task-

\begin{itemize}
    \item Level/Task 1: Classify the memory data into two groups: i) benign and ii) malware.
    
   \item Level/Task 2: Classify the malware data into 3 types: i) Trojan Horse, ii) Spyware, and ii) Ransomware.

   \item  Level/Task 3: Classify different types of malware into the number of families they represent.

\end{itemize}

\subsection{Feature Selection}
In the feature selection phase, we utilize statistical approaches to identify highly class-correlated features. To address computational constraints, we prefer filter-based statistical methods over wrapper methods, as they avoid exhaustive searches and classifier integration. Besides, we emphasize minimizing the number of selected features during this step to enhance computational efficiency and minimize security concerns in subsequent stages. Thus, our goal is to identify a feature selection method that demonstrates robust performance while utilizing only a minimal number of features.

We employ three filter based feature selection approaches, identifying the most relevant features as described below.

\subsubsection{Chi-square}
We compute the Chi-square ($\chi 2$) value between each feature and the corresponding class. Utilizing the Chi-square statistics, features that do not correlate well with class labels are eliminated as they may not be significant for classification.

The Chi-square ($\chi 2$) value is calculated as follows - 

\begin{equation}
\chi 2 = \sum {\frac {{(O_i - E_i)}^2}{E_i}}
\end{equation}

where, $O_i$ represents observed value and $E_i$ represents expected value of feature $i$. A gene with a high chi-square value is highly correlated with the class; hence, important for classification.

\subsubsection{Analysis of Variance (ANOVA)}
ANOVA is a statistical method that examines the means of two or more groups to find whether they are significantly different from each other. ANOVA F-test determines the variance between and within groups, calculates the F score, and uses it to identify informative features.

\subsubsection{Mutual Information (MI) }
Mutual information is a measure of the statistical dependency between two random variables. In the context of feature selection, it quantifies the amount of information a feature (i.e., memory-related attribute) provides about the target variable (class of the sample). A higher score indicates a stronger relationship between the feature and the target variable, indicating that the feature is potentially more informative for the prediction task.

The mutual information between a feature X and a class label Y is computed as follows-\newline

\begin{equation}
MI(X, Y) = H(X) + H(Y) - H(X, Y)
\end{equation}

where
$MI(X, Y)$ represents the mutual information between feature X and class label Y.
$H(X)$ is the entropy of feature $X$.
$H(Y)$ is the entropy of class label Y.
$H(X, Y)$ is the joint entropy of feature $X$ and class label $Y$.
Entropy measures the uncertainty or disorder in a random variable. The calculation of entropy involves probability distributions of the variables. The specific formula for entropy calculation depends on the type of data and probability distribution considered. We use the default implementation of the mutual information algorithm of scikit-learn \cite{scikit-learn}.

\begin{table*}
\caption{Performance of various classifiers for malware identification task (i.e., malicious or benign)}
\centering
\begin{tabular}{c|cc|cccccc|cccccc}
\\
\textbf{Classifier}  &  \multicolumn{2}{c}\textbf{\textbf{100\% percent}}  & \multicolumn{6}{c}{\textbf{25\% percent}} &  \multicolumn{6}{c}{\textbf{50\% percent}}
\\
\hline
& &  &  \multicolumn{2}{c} {\textbf{Chi}} & \multicolumn{2}{c} {\textbf{ANOVA}} & \multicolumn{2}{c} {\textbf{Mutual}} &    \multicolumn{2}{c} {\textbf{Chi}} & \multicolumn{2}{c} {\textbf{ANOVA}} & \multicolumn{2}{c} {\textbf{Mutual}} 
 \\

\hline
&  {F1} & {Acc.}  & {F1} & {Acc.}  & {F1} &  {Acc.} 
 &{F1} & {Acc.}  & {F1} &  {Acc.}  & {F1} &  {Acc.}  & {F1} &  {Acc.}
\\

MNB & 0.64& 0.55 &  0.64 & 0.55  & 0.98 & 0.98 & 0.98 & 0.98 & 0.64 & 0.55 & 0.98 & 0.98 & 0.64 & 0.55  \\

LDA & 0.99 &0. 99 &  0.99 & 0.99 & 0.99 & 0.99 & 0.99 & 0.99 & 0.99 & 0.99 & 0.99 & 0.99 & 0.99 & 0.99  \\

Adaboost & 1.0 & 1.0 & 1.0 & 1.0 & 1.0 & 1.0 & 1.0 & 1.0 & 1.0 & 1.0 & 1.0  & 1.0 & 1.0 & 1.0 \\ 
K-NN & 1.0 & 1.0 &   1.0 & 1.0 &  1.0 & 1.0 & 1.0 & 1.0 & 1.0 & 1.0 &  1.0 & 1. 0& 1.0 & 1.0 \\

Extra-Tree & 1.0 & 1.0 & 1.0 & 1.0 & 1.0 & 1.0 & 1.0 & 1.0 & 1.0 & 1.0 & 1.0 & 1.0 & 1.0 &  1.0 \\

Random Forest   & 1.0  & 1.0 & 1.0 & 1.0 & 1.0  & 1. 0 & 1.0 & 1.0 & 1.0 & 1.0 & 1.0 & 1.0 & 1.0 & 1.0   \\

 &  \\

\end{tabular}
\label{malware_presence_three_feature_selection}
\end{table*}

\vspace{6mm}

\begin{table*}
\caption{Performance of various classifiers with feature selection approaches for malware type classification task (3-class), bold texts represent best F1 scores for each type of feature settings.}
\centering
\begin{tabular}{c|cc|cccccc|cccccc}
\\
\textbf{Classifier}  &  \multicolumn{2}{c}\textbf{\textbf{100\% percent}}  & \multicolumn{6}{c}{\textbf{25\% percent}} &  \multicolumn{6}{c}{\textbf{50\% percent}}
\\
\hline
&  &  & \multicolumn{2}{c} {\textbf{Chi}} & \multicolumn{2}{c} {\textbf{ANOVA}} & \multicolumn{2}{c} {\textbf{Mutual}} &    \multicolumn{2}{c} {\textbf{Chi}} & \multicolumn{2}{c} {\textbf{ANOVA}} & \multicolumn{2}{c} {\textbf{Mutual}} 
 \\

\hline
&  F1 & Acc.  & F1 & Acc.  & {F1} &  {Acc.} 
 &\textbf{F1} & {Acc.}  & {F1} &  {Acc.}  & {F1} & {Acc.}  & {F1} &  {Acc.}
\\

MNB &  0.40 & 0.36 & 0.40 & 0.36 & 0.36 & 0.35 & 0.41 & 0.36 & 0.40 & 0.36 & 0.38 & 0.36 & 0.39 & 0.36 \\

LDA & 0.47 & 0.47 & 0.38 & 0.38 & 0.40 & 0.40 & 0.41 & 0.41 & 0.44 & 0.44 & 0.41 & 0.41 & 0.46 & 0.46 \\

Adaboost & 0.60 & 0.59 & 0.52 & 0.52 & 0.48 & 0.48 & 0.59 & 0.59 & 0.55 & 0.55 & 0.53 & 0.53 & 0.59 & 0.59 \\ 

K-NN & 0.62 & 0.62 & 0.58 & 0.58 & 0.48 & 0.48 & 0.69 & 0.70 & 0.60 & 0.60 & 0.57 & 0.57 & 0.63 & 0.63  \\

Extra-Tree & 0.73 & 0.73 & 0.66 & 0.66 & 0.57 & 0.57 & 0.74 & 0.74 &  0.72 & 0.72 & 0.64 & 0.64 & 0.75 & 0.75\\

Random Forest   & 0.75 & 0.75 & 0.68 & 0.68 & 0.59 & 0.59 & \textbf{0.75} & 0.75 &  0.74 & 0.74 & 0.65 & 0.65 &  \textbf{0.76} & 0.76    \\

 &  \\

\end{tabular}
\label{3_class_feature_selection}
\end{table*}

\subsection{Classification}
We employ a number of machine learning classifiers (described below) for the classification tasks with selected features. For all the classifiers, the default parameter settings of scikit-learn library \cite{scikit-learn} are used. 

\subsubsection{Random Forest (RF)}
Random Forest is an ensemble learning method that combines multiple decision trees. It is known for its robustness, ability to handle large datasets with high-dimensional features, and resistance to overfitting. Random Forest can handle both classification and regression tasks.

\subsubsection{Naive Bayes (NB)}
 Naive Bayes is a probabilistic classifier based on Bayes' theorem with the assumption of independence between features. Despite its simplifying assumption, Naive Bayes classifiers perform well in various domains, especially with large datasets.

\subsubsection{K-Nearest Neighbors (K-NN)}
 K-Nearest Neighbors (KNN) is a non-parametric classification method that determines the class of a data point by taking a majority vote from its closest neighbors. This approach is known for its ease of implementation and excellent performance, especially when dealing with small to medium-sized datasets and clear clusters.

\subsubsection{AdaBoost}
AdaBoost is an ensemble learning algorithm used for classification and regression tasks. It combines multiple weak learners, like decision trees, to create a powerful model by iteratively giving more weight to misclassified instances and adjusting the sample weights. This process focuses on challenging data points and improves overall accuracy. AdaBoost is versatile but sensitive to noisy data.

\subsubsection{Linear Discriminant Analysis (LDA)}

Linear Discriminant Analysis (LDA) can be used as a supervised dimensionality reduction technique for classification tasks. LDA aims to find a linear combination of features that maximizes class separability and can be used to project the data into a lower-dimensional space while preserving class information.

\subsubsection{Extra Trees}
Extra Trees, also known as Extremely Randomized Trees, is a powerful ensemble machine learning algorithm that constructs multiple decision trees using randomized subsets of features and thresholds. Unlike Random Forest, ExtraTrees introduces an additional layer of randomness in the feature and threshold selection process, effectively mitigating the risk of overfitting. Through the aggregation of predictions from individual trees, ExtraTrees can deliver robust and accurate results, making it well-suited for handling high-dimensional data, noisy features, and irrelevant variables.

\begin{table*}[hbt!]
\caption{Performance of top classifiers with various feature selection approaches for 5-class malware family classification tasks (bold texts represent best F1 scores for each type of feature settings.)}
\centering
\begin{tabular}{cc|cc|cccccc|cccccc}
\\
\textbf{Malware}  &  & \multicolumn{2}{c}{\textbf{100\% features}}  & \multicolumn{6}{c}{\textbf{25\% features}} &  \multicolumn{6}{c}{\textbf{50\% features}}
\\
\cline{3-16}

\textbf{Family} & \textbf{Classifier}  & &  &  \multicolumn{2}{c} {\textbf{Chi}} & \multicolumn{2}{c} {\textbf{ANOVA}} & \multicolumn{2}{c} {\textbf{MI}} &    \multicolumn{2}{c} {\textbf{Chi}} & \multicolumn{2}{c} {\textbf{ANOVA}} & \multicolumn{2}{c} {\textbf{MI}} 
 \\

  &    & {F1} &  {Acc.}  &{F1} & {Acc.}  & {F1} &  {Acc.} 
 &{F1} & {Acc.}  & {F1} &  {Acc.}  & {F1} &  {Acc.}  & {F1} &  {Acc.}
\\
        
\hline

& Adaboost & 0.54 &0.55 & 0.48 & 0.49 & 0.44& 0.44 & 0.53 & 0.54 & 0.52 &0.53 & 0.51& 0.51 & 0.55  &0.55 \\

Trozan &K-NN & 0.573& 0.58  & 0.54  & 0.54 &0.42 & 0.42 & 0.63 &0.63 &  0.57 & 0.57 & 0.51 & 0.51 & 0.58 &0.59 \\

&Extra-Tree & 0.72 &0.72  &0.65 & 0.66 & 0.50 & 0.50 & 0.73 & 0.73 & 0.72 &0.72 & 0.65 & 0.65 &  0.73 & 0.73 \\

&Random Forest  & 0.74 & 0.74 & 0.69& 0.69 & 0.54 & 0.54 & \textbf{0.73} & 0.73 & 0.74 &0.74 & 0.69 & 0.70 & \textbf{0.75} & 0.75 \\

\hline

& K-NN &  0.51 &0.50 & 0.52 & 0.50 & 0.5 & 0.49 &0.52 &0.51  & 0.51 & 0.5 & 0.51 & 0.50 &0.52 &0.51  \\

Spyware & Adaboost  &0.44 &0.43 & 0.39 & 0.40 & 0.38 & 0.38 &0.43 &0.42 & 0.41 & 0.41 & 0.45 & 0.44 &0.42& 0.42\\

 & Extra-Tree & 0.62 &0.61 &  0.65 & 0.64 & 0.57& 0.56 &0.63& 0.62& 0.63  & 0.62 & 0.64 & 0.62 & 0.64& 0.63\\

& Random Forest & \textbf{0.65}& 0.64 & 0.65 & 0.64 & 0.57 & 0.56 & 0.64 &0.62 & 0.65 & 0.64 & \textbf{0.66} & 0.65 &  0.65 & 0.63 \\

\hline

&K-NN  &0.45& 0.45 & 0.43 &  0.43 & 0.41 & 0. 41 & 0.50& 0.51  & 0.45 & 0.45 & 0.45 & 0.45 &0.44 &0.44 \\

Ransome &Adaboost &  0.40 &0.40 &  0.37 & 0.37 & 0.34& 0.33 & 0.395 &0.39 & 0.38 & 0.38 & 0.39 & 0.38 &0.40 & 0.40   \\

 &Extra-Tree & 0.55& 0.55 & 0.50 & 0.50 & 0.44 & 0.43 & 0.55 &0.55  & 0.50 &  0.50 & 0.53 & 0.53 &0.55& 0.55 \\

&Random Forest &0.56& 0.56 & 0.50 & 0.51 & 0.45 & 0.45 & \textbf{0.56} &0.56  & \textbf{0.57} & 0.57 & 0.54 & 0.54 &  0.56 & 0.56 \\

\end{tabular}
\label{malware_family_classification}
\end{table*}

\section{Results and Discussion}

\subsection{Evaluation Settings}

To evaluate and compare the performance of various classifiers, we consider 5-fold cross-validation. Cross-validation is generally considered better than a pre-defined training-testing split for model evaluation because it provides more reliable and unbiased estimates of a model's performance. We adopt two performance evaluation metrics: i) macro F1 and ii) accuracy. Since some of our classification tasks deal with a class-imbalanced dataset, the macro F1 score is a better estimator for the classifier's performance than accuracy, which weights all the classes equally.













\subsection{Performance comparison of feature selection with original setting}

Table \ref{malware_presence_three_feature_selection} compares the performance of various classifiers for malware and benign type classification employing multiple feature selection approaches. We assess their performances under three conditions: when using 25\% and 50\% of features selected by various feature selection methods, as well as in the original configuration employing 100\% of the features. As we can see, classifiers exhibit very similar performance with (e.g., 25\% and 50\%) and without (100\%) incorporating feature selection, which corroborates the effectiveness of the feature selection approaches. This binary-level malware identification task is relatively easy, as we can see that all the classifiers are capable of yielding perfect or almost perfect F1 scores. One interesting thing we observe is that for MNB, the performance improvement is dramatic when we reduce the number of features. When all the memory features are used, we obtain an F1 score of only 0.64. Leveraging feature selection approaches such as ANOVA and MI to identify the top 25\% of features can dramatically improve the performance of the classifier. 

Table \ref{3_class_feature_selection} shows the performance of various classifiers for the malware type classification. Similar to the malware identification task, we provide the performance of various classifiers in three different settings. As we can find from Table \ref{3_class_feature_selection}, among all the classifiers, RF performs best; it obtains an F1 score of around 0.75-0.76. One important thing we notice, all the tree-based classifiers perform better than other methods. Regarding the features used, we find that incorporating 50\% features improves performance for all the top classifiers, and this is true for all the feature selection approaches. When the best results are considered, we find MI feature selection is the most effective for classifying the top feature.

Table \ref{malware_family_classification} presents the performance of various classifiers for the malware family classification. For the 5-class classification problem of Trojan malware, we find RF with 50\% feature (28 features) selected by MI performs best with an F1 score of 0.75, compared to the F1 scores obtained as 0.74 using all features. For the Spyware 5-class categorization, the best F1 score of 0.66 is obtained by the RF classifier with 50\% features selected by ANOVA. MI with RF shows very similar results of 0.65. We find classifying the Ransomware family very challenging. For Ransomware, the best result is obtained by Chi-Square (50\% features) with RF, which is 0.57. Selecting only 25\% features through MI, RF performs similarly to using 100\% features.

\begin{figure*}
\includegraphics[width=0.83 \linewidth]{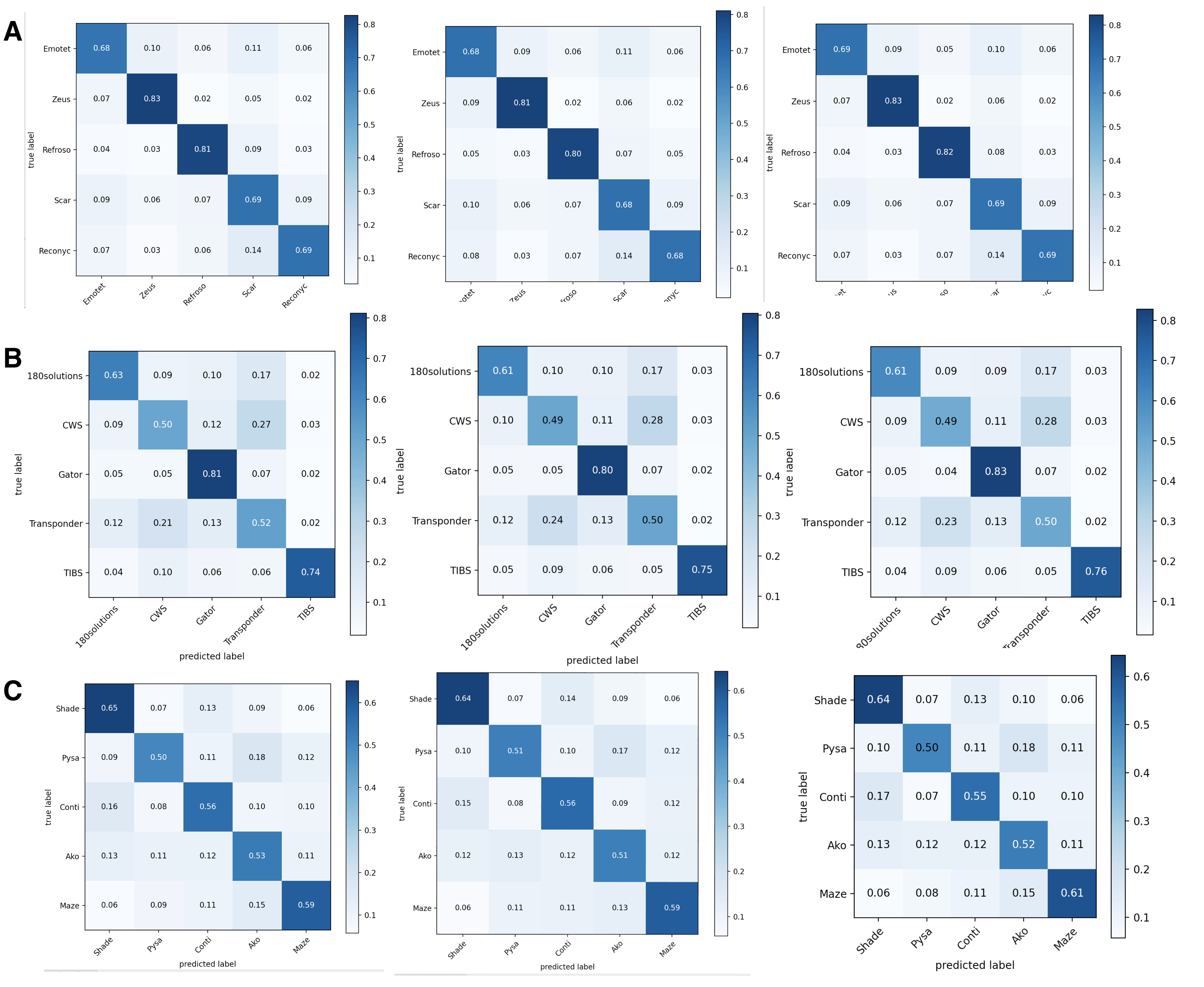}
\vspace{2.9mm}
\centering \caption{Comparison of confusion matrices MI with RF classifiers considering various numbers of features (100\% (column 1), 25\% (column 2), and 50\% (column 3)) for malware family classification (A. Trojan horse  B. Spyware  C. Ransomware)}
\label{confusion_matrix}
\end{figure*}

Based on the results of all the tasks we investigate, it is evident that we can actually reduce the number of features used for ML classifiers without compromising classifier performance. In some instances, the feature selection even improves the performance of the classifiers. For all the classification tasks (Table \ref{3_class_feature_selection} and \ref{malware_family_classification}), we notice by selecting 50\% features, we can actually obtain better results than incorporating all the features. We can reduce the percentage of features even further to 25\% without compromising efficacy. For example, when classifying the Trojan horse family, utilizing just 25\% of the features resulted in an F1 score of 0.73. Even using all features, the improvement was minimal, reaching only 0.74, a mere 1\% increase. For Spyware and Ransomware, we can actually reach the same level of performance using only 25\% of the features selected by various feature selection methods.

Among all feature selection algorithms, we observe that the MI delivers the most consistent performance. For all the tasks, in 25\% and 50\% feature settings, MI yields a similar performance of 100\% input features. Although in a few cases, we find Chi-square and ANOVA perform a bit better than MI, their performances are not consistent; In contrast, MI performance is always consistent and similar to using 100\% of features.

We further investigate whether the reduced feature sets impact the prediction patterns of the classifiers. For the highly challenging classification task-3 (malware family classifications), we show the confusion matrix of RF classifiers with 100\%, 25\%, and 50\% features (Fig \ref{confusion_matrix}). As we can see, in all three malware family classification tasks, different feature sizes show similar patterns in misclassification. The accuracy of different classes is similar in all three settings. Besides, the misclassification (prediction) also shows similar patterns for all the cases.

Regarding privacy, while feature selection may not directly address data privacy concerns, it can have an indirect impact on privacy in certain scenarios. By reducing the number of features, feature selection can potentially limit the exposure of sensitive or identifiable information during model training or analysis. This reduction in the feature space may help minimize the risk of inadvertently revealing private information.

\subsection{Comparison with other studies}

Note that to the best of our knowledge, only two studies \cite{carrier2022detecting,dener2022malware} have utilized the MalMemAnalysis-2022 dataset for malware analysis, as the dataset is relatively recent. However, both studies focused solely on the malware identification task (Task 1 of this study). The study by \cite{carrier2022detecting} achieved the best results of approximately 0.99 by employing an ensemble of NB, RF, and DT classifiers. This number is slightly lower than our feature selection strategy with the RF classifier. Unfortunately, the authors did not mention their criteria for selecting the training and testing datasets, making a direct comparison with our method difficult. On the other hand, the other study by \cite{dener2022malware} achieved F1 scores ranging from 0.99 to 1.0 using various classifiers, which is similar to our results. They used a 70\% and 30\% training and testing split; however, they did not provide details on how they selected the training and testing data, hindering a precise comparison.

Despite the missing information, all the works, including ours, demonstrate F1 scores and accuracy close to perfection, suggesting that classifying benign and malware samples in this dataset is relatively trivial.

\section{Summary and Future Work}
With the growth of Advanced Persistent Threats (APTs), more sophisticated malware has been exhibited in recent years. As malware developers continually refine their techniques and develop novel strategies to evade security controls, the need for advanced malware analysis frameworks becomes crucial. Thus, this research aims to enhance the performance of malware detection and classification tasks by incorporating feature selection approaches. We assess the performance and generalization capabilities of the various feature selection approaches for five classification tasks (one malware detection task, one malware type classification task, and three malware family classifications) through a set of experiments. The evaluation on the MalMemAnalysis-2022 dataset shows the effectiveness of the feature selection approaches (i.e., Chi-Square, ANOVA, and MI) for various scenarios. Among the three feature selection approaches, MI exhibits the best consistency and efficacy and obtains similar or better performance than the original 100\% input features by leveraging only 25\%-50\% features with the RF classifier. The reduced feature set obtained by MI  can improve the efficiency of ML classifiers, as well as can enhance privacy through data minimization as fewer data are fed to ML classifiers for the prediction task.  By selecting robust and relevant features, models can become more resilient against such attacks. Removing noisy or misleading features can improve the model's ability to generalize and reduce its susceptibility to manipulation. However, it is important to note that relying solely on feature selection is not enough to ensure comprehensive data privacy. Adopting a holistic approach that incorporates both feature selection and privacy-preserving practices (e.g., data anonymization, encryption, access controls) is essential to effectively mitigate privacy risks.

Future work will focus on expanding and refining the findings of this study. We will investigate the performance of feature selection approaches on a broader range of memory datasets beyond the MalMemAnalysis-2022. Besides, we will explore other privacy-preserving practices, such as differential privacy techniques or federated learning, to ensure a more robust and comprehensive safeguard against privacy risks. Investigating the trade-offs between model accuracy and privacy preservation, as well as areas that require attention, which we plan to study in our future works.

\bibliography{anthology}
\bibliographystyle{ieeetr}
\end{document}